\documentclass[aps,prd,preprint,superscriptaddress,groupedaddress,showpacs,amsmath,amssymb]{revtex4}
\usepackage[dvips]{graphicx}
\usepackage{amsmath,amssymb,times}

\newcommand{\bequ}{\begin{equation}}
\newcommand{\eequ}{\end{equation}}
\newcommand{\bea}{\begin{eqnarray}}
\newcommand{\eea}{\end{eqnarray}}

\DeclareSymbolFont{boldletters}{OML}{cmm} {b}{it}
\DeclareSymbolFontAlphabet{\mathbit}{boldletters}
\DeclareMathSymbol{\alpha}{\mathalpha}{letters}{"0B}
\DeclareMathSymbol{\beta}{\mathalpha}{letters}{"0C}
\DeclareMathSymbol{\gamma}{\mathalpha}{letters}{"0D}
\DeclareMathSymbol{\delta}{\mathalpha}{letters}{"0E}
\DeclareMathSymbol{\epsilon}{\mathalpha}{letters}{"0F}
\DeclareMathSymbol{\zeta}{\mathalpha}{letters}{"10}
\DeclareMathSymbol{\eta}{\mathalpha}{letters}{"11}
\DeclareMathSymbol{\theta}{\mathalpha}{letters}{"12}
\DeclareMathSymbol{\iota}{\mathalpha}{letters}{"13}
\DeclareMathSymbol{\kappa}{\mathalpha}{letters}{"14}
\DeclareMathSymbol{\lambda}{\mathalpha}{letters}{"15}
\DeclareMathSymbol{\mu}{\mathalpha}{letters}{"16}
\DeclareMathSymbol{\nu}{\mathalpha}{letters}{"17}
\DeclareMathSymbol{\xi}{\mathalpha}{letters}{"18}
\DeclareMathSymbol{\pi}{\mathalpha}{letters}{"19}
\DeclareMathSymbol{\rho}{\mathalpha}{letters}{"1A}
\DeclareMathSymbol{\sigma}{\mathalpha}{letters}{"1B}
\DeclareMathSymbol{\tau}{\mathalpha}{letters}{"1C}
\DeclareMathSymbol{\upsilon}{\mathalpha}{letters}{"1D}
\DeclareMathSymbol{\phi}{\mathalpha}{letters}{"1E}
\DeclareMathSymbol{\chi}{\mathalpha}{letters}{"1F}
\DeclareMathSymbol{\psi}{\mathalpha}{letters}{"20}
\DeclareMathSymbol{\omega}{\mathalpha}{letters}{"21}
\DeclareMathSymbol{\varepsilon}{\mathalpha}{letters}{"22}
\DeclareMathSymbol{\vartheta}{\mathalpha}{letters}{"23}
\DeclareMathSymbol{\varpi}{\mathalpha}{letters}{"24}
\DeclareMathSymbol{\varrho}{\mathalpha}{letters}{"25}
\DeclareMathSymbol{\varsigma}{\mathalpha}{letters}{"26}
\DeclareMathSymbol{\varphi}{\mathalpha}{letters}{"27}
\DeclareMathSymbol{\Gamma}{\mathalpha}{letters}{"00}
\DeclareMathSymbol{\Delta}{\mathalpha}{letters}{"01}
\DeclareMathSymbol{\Theta}{\mathalpha}{letters}{"02}
\DeclareMathSymbol{\Lambda}{\mathalpha}{letters}{"03}
\DeclareMathSymbol{\Xi}{\mathalpha}{letters}{"04}
\DeclareMathSymbol{\Pi}{\mathalpha}{letters}{"05}
\DeclareMathSymbol{\Sigma}{\mathalpha}{letters}{"06}
\DeclareMathSymbol{\Upsilon}{\mathalpha}{letters}{"07}
\DeclareMathSymbol{\Phi}{\mathalpha}{letters}{"08}
\DeclareMathSymbol{\Psi}{\mathalpha}{letters}{"09}
\DeclareMathSymbol{\Omega}{\mathalpha}{letters}{"0A}

\begin{document}
\preprint{SAGA-HE-277,  RIKEN-QHP-62}

\title{Confinement and $\mathbb{Z}_{3}$ symmetry in three-flavor QCD}

\author{Hiroaki Kouno}
\email[]{kounoh@cc.saga-u.ac.jp}
\affiliation{Department of Physics, Saga University,
             Saga 840-8502, Japan}

\author{Takahiro Makiyama}
\email[]{12634019@edu.cc.saga-u.ac.jp}
\affiliation{Department of Physics, Saga University,
             Saga 840-8502, Japan}

\author{\\Takahiro Sasaki}
\email[]{sasaki@phys.kyushu-u.ac.jp}
\affiliation{Department of Physics, Graduate School of Sciences, Kyushu University,
             Fukuoka 812-8581, Japan}

\author{Yuji Sakai}
\email[]{ysakai@riken.jp}
\affiliation{Theoretical Research Division, Nishina Center, RIKEN, Saitama 351-0198, Japan}

\author{Masanobu Yahiro}
\email[]{yahiro@phys.kyushu-u.ac.jp}
\affiliation{Department of Physics, Graduate School of Sciences, Kyushu University,
             Fukuoka 812-8581, Japan}

\date{\today}

\begin{abstract}
We investigate the confinement mechanism in three-flavor QCD with 
imaginary isospin chemical potentials 
$(\mu_u,\mu_d,\mu_s)=(i\theta T,-i\theta T,0)$, using 
the Polyakov-loop extended Nambu--Jona-Lasinio (PNJL) model, where 
$T$ is temperature. 
As for three degenerate flavors, the system has $\mathbb{Z}_{3}$ symmetry 
at $\theta=2\pi/3$ and hence the Polyakov loop $\Phi$ vanishes there 
for small $T$. As for 2+1 flavors, the symmetry is not preserved 
for any $\theta$, but $\Phi$ becomes zero 
at $\theta=\theta_{\rm conf} < 2\pi/3$ for small $T$. 
The confinement phase defined by 
$\Phi=0$ is realized, even if the system does not have 
$\mathbb{Z}_{3}$ symmetry exactly. 
In the $\theta$-$T$ plane, there is a critical endpoint 
of deconfinement transition. 
The deconfinement crossover at zero chemical potential 
is a remnant of the first-order deconfinement transition 
at $\theta=\theta_{\rm conf}$. 
The relation between the non-diagonal element $\chi_{us}$ of 
quark number susceptibilities and 
the deconfinement transition is studied. 
The present results can be checked by lattice QCD simulations directly, 
since the simulations are free from the sign problem for any $\theta$. 
\end{abstract}

\maketitle

\section{Introduction}
\label{Introduction}
 
Lattice QCD (LQCD) simulations indicate that QCD is in the confinement 
and chiral symmetry breaking phase at low temperature ($T$) and 
in the deconfinement and chiral symmetry restoration phase at high $T$. 
Understanding of the confinement mechanism is, nevertheless, not adequate, 
although the chiral restoration is relatively well understood. 
The reason mainly comes from the fact that there is no exact symmetry 
for the deconfinement transition 
and hence the order parameter is unknown. 
In the limit of infinite current quark mass, 
the Polyakov-loop~\cite{Polyakov} is an exact order parameter for the deconfinement transition, since $\mathbb{Z}_{N_c}$ symmetry is exact there, 
where $N_c$ is the number of colors. 
The chiral condensate is, meanwhile, an exact order parameter 
for the chiral restoration in the limit of zero current quark mass. 
In the real world where $u$ and $d$ quarks have small current masses 
$m_l \equiv m_u=m_d$, 
the chiral condensate is considered to be a good order parameter 
for the chiral restoration, but there is no guarantee 
that the Polyakov-loop $\Phi$ is a good order parameter for 
the deconfinement transition. 

In order to answer this problem, 
we constructed a gauge theory invariant 
under the $\mathbb{Z}_{N_c}$ transformation, that is, a gauge theory 
with $N_f$ degenerate flavor fermions having flavor-dependent 
fermion boundary conditions~\cite{Kouno_TBC,Sakai_TBC}. 
This $\mathbb{Z}_{N_c}$-symmetric gauge theory is constructed as follow. 
Let us start with $N_c$-color QCD. The partition function $Z$ in 
Euclidean spacetime is   
\bea
Z=\int Dq D\bar{q} DA \exp[-S_0] 
\label{QCD-Z}
\eea
with the action
\bea
S_0=\int d^4x [\sum_{f}\bar{q}_f(\gamma_\nu D_\nu +m_f)q_f
+{1\over{4g^2}}{F_{\mu\nu}^{a}}^2], 
\label{QCD-S}
\eea
where $q_f$ is the quark field with flavor $f$ and current quark mass $m_f$, 
$D_\nu =\partial_\nu-iA_\nu$ is the covariant derivative 
with the gauge field $A_\nu$, $g$ is the gauge coupling and  $F_{\mu\nu}=\partial_\mu A_\nu -\partial_\nu A_\mu -i[A_\mu ,A_\nu ]=F_{\mu\nu}^aT^a$ 
for the SU($N_c$) generators $T^a$.   
The temporal boundary conditions for quarks are 
\bea
q_f(x, \beta=1/T )=-q_f(x, 0). 
\label{period-QCD}
\eea
The boundary conditions are changed into 
\bea
q_f(x, \beta)=-\exp{(i 2\pi k/{N_c})}q_f(x, 0)  
\label{period-QCD-Z}
\eea
by the $\mathbb{Z}_{N_c}$ transformation~~\cite{Polyakov,Susskind,MS_ZN,RW,Philipsen_review} 
\begin{eqnarray}
q &\to& 
Uq,
\nonumber\\
A_\nu &\to& 
UA_\nu U^{-1}-i(\partial_\nu U)U^{-1},
\label{ZNtrans}
\end{eqnarray}
where $U(x,\tau )$ are elements of SU$(N_c)$ with the property $U(x,\beta )=\exp{(-i2\pi k/N_c)}U(x,0)$ for integer $k$, 
while the action $S_0$ keeps the original form \eqref{QCD-S} 
since $\mathbb{Z}_{N_c}$ symmetry is 
the center symmetry of the gauge symmetry~\cite{RW}. 
$\mathbb{Z}_{N_c}$ symmetry thus breaks down 
through the fermion boundary condition in QCD. 

Now we consider the SU($N$) gauge theory with $N$ degenerate flavor 
quarks, i.e. $N \equiv N_f=N_c$, and assume 
the following flavor dependent twist boundary conditions (TBC): 
\begin{eqnarray}
q_f(x, \beta )&=&-\exp{(-i\theta_f)}q_f(x,0)
\nonumber\\
&\equiv& -\exp{[-i(\theta_1+2\pi (f-1)/N)]}q_f(x, 0)
\label{period}
\end{eqnarray}
for flavors $f$ labeled by integers from 1 to $N$; 
here $\theta_1$ is an arbitrary real number in a range 
of $0 \le \theta_1 < 2\pi$. 
The action $S_0$ with the TBC is invariant under the $\mathbb{Z}_{N_c}$ 
transformation. 
In fact, the $\mathbb{Z}_{N_c}$ transformation changes $f$ into $f-k$, 
but $f-k$ can be relabeled by $f$ since $S_0$ is invariant under 
the relabeling. This is the gauge theory proposed in our previous 
works~\cite{Kouno_TBC,Sakai_TBC} and is referred to as 
the $\mathbb{Z}_{N_c}$-symmetric gauge theory in this paper. 
.

When the fermion field $q_f$ is replaced by 
\begin{eqnarray}
q_f \to \exp{(-i\theta_fT\tau )}q_f 
\label{transform_1}
\end{eqnarray}
with Euclidean time $\tau$, 
the action $S_0$ is changed into~\cite{Philipsen_review}
\bea
S(\theta_f)=\int d^4x [\sum_{f}\bar{q}_f
(\gamma_\nu D_\nu - \mu_f\gamma_4+m_f)q_f
+{1\over{4g^2}}F_{\mu\nu}^2] 
\notag \\
\label{QCD1}
\eea
with the imaginary quark number chemical potential $\mu_f=i T \theta_f$, 
while  
the TBC is transformed back to the standard one \eqref{period-QCD}. 
The action $S_0$ with the TBC is thus equivalent to 
the action $S(\theta_f)$ with the standard one \eqref{period-QCD}. 

In the limit of $T=0$, the $\mathbb{Z}_{N_c}$-symmetric 
gauge theory is identical with QCD with the standard boundary 
condition \eqref{period-QCD}. Noting that $\Phi=0$ at $T=0$, 
one can predict that in the $\mathbb{Z}_{N_c}$-symmetric 
gauge theory 
$\mathbb{Z}_{N_c}$ symmetry is preserved up to some temperature $T_c$ 
and spontaneously broken above $T_c$. 
In fact, this behavior is confirmed in Refs.~\cite{Kouno_TBC,Sakai_TBC} 
by imposing the TBC on the Polyakov-loop extended Nambu-Jona-Lasinio (PNJL) model~\cite{Meisinger,Dumitru,Fukushima,Ratti,Rossner,Schaefer,Abuki,Fukushima2,Kashiwa1,Sakai,McLerran_largeNc,Sakai2,Kashiwa5,Kouno,Hell,Sakai_imiso,Matsumoto,Sasaki-T,Sakai5,Gatto,Sasaki-T_Nf3,BL,Sakai_hadron,Kashiwa_NL_IM,Morita,Sakai_CP,Bhattacharyya,Pagura_im,Megias,Lour} that has the same global symmetries as QCD at imaginary 
chemical potentials. 
In the $\mathbb{Z}_{N_c}$-symmetric gauge theory, the quarkyonic phase with 
$\Phi=0$ and finite quark-number density appear at small $T$ and large 
real quark-number chemical potential $\mu$~\cite{Sakai_TBC}. 
$\mathbb{Z}_{N_c}$ symmetry is thus essential for emergence of 
the quarkyonic phase.

\begin{figure}[htbp]
\begin{center}
\vspace{0.5cm}
\includegraphics[width=0.2\textwidth]{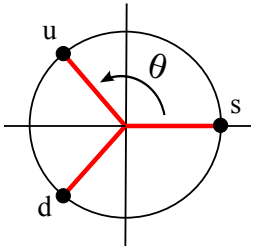}
\end{center}
\vspace{-10pt}
\caption{The $\theta$-variant TBC: Location of $e^{i\theta_f}$ ($f=u,d,s$) 
on a unit circle in the complex plane. 
}
\label{theta_f}
\end{figure}

In real QCD, however, 
s-quark has a current quark mass $m_s$ heavier than $m_l$, where $m_l$ is a current quark mass of light quarks.  
This means that real QCD does not become $\mathbb{Z}_{3}$-symmetric 
even if the TBC is imposed. 
However, the TBC or its small extension may be important 
to understand the confinement mechanism, since 
real QCD with 2+1 flavors is not far from QCD with three degenerate flavors. 
As an extension of the TBC, one can consider the $\theta$-variant TBC
\bea
(\theta_u,\theta_d,\theta_s)=(\theta,-\theta,0), 
\label{theta-variant TBC}
\eea
where $\theta$ varies from 0 to $\pi$; this boundary condition is 
illustrated in Fig.~\ref{theta_f}. 
The system with the $\theta$-variant TBC is equivalent to the system 
with imaginary isospin chemical potential 
\bea
(\mu_u,\mu_d,\mu_s)=(i\theta T,-i\theta T,0), 
\label{imaginary isospin-chemical potentials}
\eea
where 
$\mu_{\rm I} \equiv \mu_u-\mu_d=2i\theta T$ means 
the imaginary isospin chemical potential. 
The $\theta$-variant TBC agrees with the standard boundary condition 
\eqref{period-QCD} when $\theta=0$ and the TBC \eqref{period} 
when $\theta=2\pi/3$. 
For the case of small $T$, as shown later in Sec.~\ref{results}, 
$\Phi$ keeps a real number for any $\theta$ and 
varies from a positive value to a negative one as $\theta$ increases 
from 0, 
since charge-conjugation (${\cal C}$) symmetry 
is not spontaneously broken there. 
This means that $\Phi$ becomes zero at some value $\theta_{\rm conf}$ 
of $\theta$. 
Hence the static-quark free energy, $-T \ln[\Phi]$, 
diverges there. 
Thus the confinement appears, 
even if the system does not have $\mathbb{Z}_{N_c}$ symmetry exactly. 
This fact means that one can consider the confinement by using $\Phi$ 
and regard it as an order parameter of the confinement/deconfinement 
transition, even if $\mathbb{Z}_{N_c}$ symmetry is not preserved exactly. 
Thus 2+1 flavor QCD with imaginary isospin-chemical potentials 
\eqref{imaginary isospin-chemical potentials} is an important system 
to be analyzed.

As for zero and real $\mu$, meanwhile, 
an important question is what is a good indicator of QCD phase transition 
in heavy-ion collisions. About 10 years ago, it was proposed 
that correlations and fluctuations 
(or susceptibilities) of conserved charges 
may be good signals~\cite{SRS,AHM,JK,Fujii,HI,KMR}.  
If there exists a critical endpoint (CEP) of chiral 
transition~\cite{AY,Barducci_CEP}, the baryon number susceptibility 
may be a good signal, since it diverges at the CEP~\cite{SRS,HI,Fujii}. 
More recently it was suggested that third moments of conserved charges 
can serve as probes of the CEP~\cite{AEK}.  
Furthermore, it was proposed in Ref.~\cite{KMR} 
with the two-phase description 
of quark-hadron phase transition that 
the non-diagonal element $\chi_{us}$ of quark number susceptibilities  
may be a good indicator of the confinement/deconfinement transition, 
since it vanishes in the deconfinement phase 
where interactions are weak but becomes finite in the hadron phase 
where different species of quarks are confined. 
The non-diagonal element was recently analyzed 
by LQCD simulations~\cite{BFKKRS,HotQCD} and 
the PNJL model~\cite{Cristoforetti,Ratti_sus}.

In this paper, we analyze QCD at imaginary isospin-chemical potentials 
\eqref{imaginary isospin-chemical potentials} for both cases of 
three degenerate flavors and 2+1 flavors, using the PNJL model. 
As for three degenerate flavors, 
QCD with imaginary isospin-chemical potentials 
\eqref{imaginary isospin-chemical potentials} becomes 
$\mathbb{Z}_{3}$-symmetric at $\theta=2\pi/3$ and hence $\Phi$ becomes zero 
there. As for 2+1 flavor quarks, $\Phi$ vanishes at 
$\theta =\theta_{\rm conf} < 2\pi/3$ and $\theta_{\rm conf}$ is a function of 
$T$. In the $\theta$-$T$ plane, there is a line of $\Phi=0$ for 
both three degenerate flavors and 2+1 flavors. 
On the line, the thermodynamic potential $\Omega$ has 
the same property between the two cases. 
For real QCD with 2+1 flavors, it is well known that 
the deconfinement transition is crossover at zero $\theta$~\cite{YAoki_nature}. 
The deconfinement crossover is a remnant of 
the first-order deconfinement transition at $\theta =\theta_{\rm conf}$. 
This $\theta$ dependence indicates that 
there exists a CEP of deconfinement transition in the $\theta$-$T$ plane. 

We also investigate the interplay between the deconfinement transition 
and the non-diagonal element $\chi_{us}$ of quark number susceptibilities 
at zero and finite $\theta$.  The pseudocritical temperature 
of the deconfinement transition is usually defined by the peak position of 
the Polyakov-loop susceptibility.  
The absolute value $|d \chi_{us}/dT|$ has a peak at the pseudocritical 
temperature, whereas $\chi_{us}$ does not. 
This behavior is more conspicuous at the CEP than at zero $\theta$. 
In the $\theta$-$T$ plane, hence, a transition line defined 
by the peak position of $|d \chi_{us}/dT|$ almost coincides 
with a transition line by the peak position of the Polyakov-loop susceptibility. 

This paper is organized as follows. 
In \S 2, the three-flavor PNJL model is recapitulated and 
$\theta$ dependence of $\Phi$ is analyzed with the PNJL model. 
In \S 3, numerical results are shown. 
Section 4 is devoted to a summary.

\section{PNJL model}
\label{PNJL model}

The PNJL model~\cite{Meisinger,Dumitru,Fukushima,Ratti,Rossner,Schaefer,Abuki,Fukushima2,Kashiwa1,Sakai,McLerran_largeNc,Sakai2,Kashiwa5,Kouno,Hell,Sakai_imiso,Matsumoto,Sasaki-T,Sakai5,Gatto,Sasaki-T_Nf3,BL,Sakai_hadron,Kashiwa_NL_IM,Morita,Sakai_CP,Bhattacharyya,Pagura_im} is designed to describe the 
confinement mechanism as well as the chiral symmetry breaking. 
In the imaginary $\mu$ region, the model can reproduce LQCD data~\cite{FK_im,FP,D'Elia,Chen34,Chen,D'Elia-iso,Cea,D'Elia-3,FP2010,Nagata,Takaishi}, since the model 
describes the Roberge-Weiss (RW) periodicity~\cite{RW,Philipsen_review,Sakai,Kouno,Morita,Pagura_im}. 
The model is also successful for the imaginary isospin chemical potential 
region~\cite{D'Elia-iso,Cea,Sakai_imiso}. We recapitulate the model.

The three-flavor PNJL Lagrangian is defined in Euclidian spacetime as
\begin{align}
{\cal L}&={\bar q}(\gamma_\nu D_\nu+{\hat m}-\hat{\mu}\gamma_4)q  
-G_{\rm S}
\sum_{a=0}^{8}[({\bar q}\lambda_a q)^2+({\bar q}i\gamma_5\lambda_a q )^2] 
\nonumber\\
&+G_{\rm D}\left[\det_{ij}{\bar q}_i(1+\gamma_5)q_j+{\rm h.c.}\right]
+{\cal U}(\Phi [A],\Phi^*[A],T),
\label{L_nc3}
\end{align} 
where $D_\nu=\partial_\nu-i\delta_{\nu4}A_4$, $\lambda_a$ is the Gell-Mann 
matrices, ${\hat m}={\rm diag}(m_u,m_d,m_s)$ denotes the mass matrix and 
${\hat \mu}={\rm diag}(\mu_u,\mu_d,\mu_s)$ stands for the chemical potential 
matrix. 
$G_{\rm S}$ and $G_{\rm D}$ are coupling constants 
of the scalar-type four-quark and the Kobayashi-Maskawa-'t Hooft (KMT) 
interaction~\cite{KMK,tHooft}, respectively. 
The KMT interaction breaks $U_\mathrm{A} (1)$ symmetry explicitly. 
The Polyakov-loop $\Phi$ and its conjugate $\Phi^*$~\cite{Polyakov,Megias_PRL} are defined by
\begin{align}
\Phi &= {1\over{3}}{\rm tr}_c(L),\quad
\Phi^* ={1\over{3}}{\rm tr}_c({\bar L}),
\label{Polyakov_nc3}
\end{align}
with $L=\exp(i A_4/T)$ and its hermitian conjugate ${\bar L}$. 
In the PNJL model, they are calculated with the Polyakov gauge. 
We take the Polyakov potential of Ref.~\cite{Rossner}:
\begin{align}
&{\cal U} = T^4 \Bigl[-\frac{a(T)}{2} {\Phi}^*\Phi\notag\\
      &~~~~~+ b(T)\ln(1 - 6{\Phi\Phi^*}  + 4(\Phi^3+{\Phi^*}^3)
            - 3(\Phi\Phi^*)^2 )\Bigr] ,
            \label{eq:E13}\\
&a(T)   = a_0 + a_1\Bigl(\frac{T_0}{T}\Bigr)
                 + a_2\Bigl(\frac{T_0}{T}\Bigr)^2,~~~~
b(T)=b_3\Bigl(\frac{T_0}{T}\Bigr)^3 .
            \label{eq:E14}
\end{align}
Parameters of $\mathcal{U}$ are fitted to LQCD data 
at finite $T$ in the pure gauge limit. 
The parameters except $T_0$ are summarized in Table \ref{table-para}.  
The Polyakov potential yields the first-order deconfinement phase transition 
at $T=T_0$ in the pure gauge theory~\cite{Boyd,Kaczmarek}. 
The original value of $T_0$ is $270$ MeV determined from the pure gauge 
LQCD data, but the PNJL model with this value yields a larger 
value of the pseudocritical temperature $T_\mathrm{c}$ 
at zero chemical potential than $T_c\approx 160$~MeV predicted 
by full LQCD \cite{Borsanyi,Soeldner,Kanaya,Bazavov}. 
We then rescale $T_0$ to 195~MeV so as to reproduce 
$T_c\sim 160$~MeV~\cite{Sasaki-T_Nf3}. 
\begin{table}[h]
\begin{center}
\begin{tabular}{llllll}
\hline \hline
~~~~~$a_0$~~~~~&~~~~~$a_1$~~~~~&~~~~~$a_2$~~~~~&~~~~~$b_3$~~~~~
\\
\hline
~~~~3.51 &~~~~-2.47 &~~~~15.2 &~~~~-1.75\\
\hline \hline
\end{tabular}
\caption{
Summary of the parameter set in the Polyakov-potential sector 
determined in Ref.~\cite{Rossner}. 
All parameters are dimensionless. 
}
\label{table-para}
\end{center}
\end{table}

The thermodynamic potential 
(per volume) is obtained by the mean-field approximation as~\cite{Matsumoto}  
\begin{align}
\Omega
&=-2\sum_{f=u,d,s}\int\frac{d^3{\bf p}}{(2\pi)^3}
\Bigl[3E_f 
+\frac{1}{\beta}\left(\ln{\cal F}_f+\ln{\cal F}_{\bar f}\right)
\nonumber\\
&+U_{\rm M}(\sigma_f)+{\cal U}(\Phi,T), 
\label{PNJL-Omega}
\end{align}
where 
\begin{eqnarray}
{\cal F}_f=&1+3\Phi e^{-\beta E^-_f}+3\Phi^* e^{-2\beta E^-_f}
+e^{-3\beta E^-_f},\label{factor_3_omega_1}\\
{\cal F}_{\bar f}=&1+3\Phi^* e^{-\beta E^+_f}+3\Phi e^{-2\beta E^+_f}
+e^{-3\beta E^+_f} 
\label{factor_3_omega_2}
\end{eqnarray}
with $\sigma_{f}=\langle{\bar q}_fq_f\rangle$, $E^{\pm}_f=E_f\pm\mu_f$
and $E_f=\sqrt{{\bf p}^2+{M_{f}}^2}$. 
The term ${\cal F}_f$ (${\cal F}_{\bar f}$) comes 
from a quark (antiquark) loop. 
For imaginary chemical potential, 
$\Phi^*$ is the complex conjugate to $\Phi$, since $\Omega$ is real. 
For imaginary isospin chemical potential, $\Omega$ is invariant under 
the ${\cal C}$ transformation $\Phi \leftrightarrow \Phi^*$, because 
\bea
\Omega (\theta) \xrightarrow{\cal C} \Omega (-\theta) 
\xrightarrow{u \leftrightarrow d } \Omega (\theta), 
\eea
where the second transformation is the relabeling of u and d. 
The three-dimensional cutoff is taken for the momentum integration 
in the vacuum term~\cite{Matsumoto}.
The dynamical quark masses $M_{f}$ and the mesonic potential $U_{\rm M}$ 
are defined by 
\begin{align}
M_{f}&=m_{f}-4G_{\rm S}\sigma_{f}+
2G_{\rm D}\sigma_{f^\prime }\sigma_{f^{\prime\prime}}, \\
U_{\rm M}&=\sum_{f=u,d,s}  2 G_{\rm S} \sigma_{f}^2 
-4 G_{\rm D} \sigma_{u}\sigma_{d}\sigma_{s}, 
\end{align}
where $f\neq f^\prime$, $f\neq f^{\prime\prime}$ and $f^\prime \neq f^{\prime\prime}$.

The NJL sector of the PNJL model has six parameters, 
($G_{\rm S}$, $G_{\rm D}$, $m_u$, $m_d$, $m_s$, $\Lambda$). 
A typical set of the parameters is obtained in Ref.~\cite{Rehberg}; 
for example, $m_l=m_u=m_d=5.5$~MeV and $m_s=140.7$~MeV. 
The parameter set is fitted to empirical values of 
$\eta'$- and $\pi$-meson masses and 
$\pi$-meson decay constant at vacuum. 
We refer to this parameter set as ``set R" (realistic parameter set). 
For theoretical interest, furthermore, we vary the strange quark mass only 
as $m_s=5.5$ and $600$~MeV. 
We refer to the parameter set with $m_s=5.5$~MeV as 
``set S" (symmetric parameter set) and that with $m_s=600$~MeV as 
``set H" (heavy strange quark parameter set). 
These parameter sets are summarized in Table~\ref{Table_NJL}.

\begin{table}[h]
\begin{center}
\begin{tabular}{lllllll}
\hline
Set &~~$m_l(\rm MeV)$~~&~~$m_s(\rm MeV)$~~&~~$\Lambda(\rm MeV)$~~~&~~~$G_{\rm S} \Lambda^2$
~~&~~~$G_{\rm D} \Lambda^5$~~
\\
\hline
R &~~~~~~5.5 &~~~~~~140.7 &~~~~~602.3 &~~~1.835 &~~~12.36 &~~~~\\
\hline
S &~~~~~~5.5 &~~~~~~5.5~~~ &~~~~~602.3 &~~~1.835 &~~~12.36 &~~~~\\
\hline
H &~~~~~~5.5 &~~~~~~600.0 &~~~~~602.3 &~~~1.835 &~~~12.36 &~~~~\\
\hline
\end{tabular}
\caption{
Summary of the parameter sets in the NJL sector. 
\label{Table_NJL}
}
\end{center}
\end{table}

The resultant thermodynamic potential $\Omega$ is a function of 
$\sigma_u$, $\sigma_d$, $\sigma_s$, $\Phi$ and $\Phi^*$. 
Values of the mean fields are determined from the location of the 
global minimum of $\Omega$ in the variable space.    
Since $\sigma_u=\sigma_d$ is preserved except for 
the RW phase in which ${\cal C}$ symmetry is spontaneously 
broken~\cite{Kouno}, 
we use $\sigma \equiv (\sigma_u+\sigma_d+\sigma_s)/3$ and 
$\sigma^\prime =\sigma_s-(\sigma_u+\sigma_d)/2=\sigma_u-\sigma_d$ 
as order parameters, and adopt 
$\Phi_{\rm R}=(\Phi +\Phi^*)/2$ and $\Phi_{\rm I}=(\Phi -\Phi^*)/(2i)$ 
instead of $\Phi$ and $\Phi^*$. 
The order parameter $\Phi_{\rm I}$ is ${\cal C}$-odd, whereas 
$\sigma$, $\sigma^\prime$ and $\Phi_{\rm R}$ are ${\cal C}$-even. 

The susceptibilities are calculable as \cite{Kashiwa1} 
\begin{eqnarray}
\chi_{\varphi_i\varphi_j}\equiv (C^{-1})_{\varphi_i\varphi_j} 
\label{sus}
\end{eqnarray}
with the curvature matrix $C$ defined by 
\begin{eqnarray}
C_{\varphi_i\varphi_j}={\partial^2 \Omega (T,\mu ,\varphi )\over{\partial \varphi_i\partial \varphi_j}},
\label{curve}
\end{eqnarray}
where $\varphi_1=\sigma$, $\varphi_2=\sigma^\prime$, $\varphi_3=\Phi_{\rm R}$, 
$\varphi_4=\Phi_{\rm I}$. When it is not confusing, we denote 
the set $\{\mu_u,\mu_d,\mu_s\}$ by $\mu$ 
and the set $\{\sigma,\sigma^\prime,\Phi_{\rm R},\Phi_{\rm I}\}$ by $\varphi$. 
The Polyakov-loop susceptibilities, 
$\chi_{\Phi_{\rm R}\Phi_{\rm R}}$ and $\chi_{\Phi_{\rm I}\Phi_{\rm I}}$ are real for imaginary isospin chemical potential 
except for the RW-phase.   

The susceptibilities of quark number densities are defined as 
\begin{eqnarray}
\chi_{ff^\prime }=-{D^2 \Omega (T,\mu, \varphi (T,\mu ))\over{D \mu_f D \mu_{f^\prime}}}, 
\label{numbersus}
\end{eqnarray}
where $f,f^\prime =u,d,s$ and the derivation ${D\over{D\mu_f}}$ means 
the partial derivation with respect to $\mu_f$ with fixing the other 
external parameters $T$ and $\mu_{f^{\prime\prime}}(\neq \mu_f)$. 
We also define the derivative of the susceptibilities with respect to $T$ as 
\begin{eqnarray}
\chi_{ff^\prime, T}=-{D \chi_{ff^\prime} (T,\mu, \varphi (T,\mu ))\over{D T}}. 
\label{numbersus_T}
\end{eqnarray}

Using the ${\mathbb Z}_3$ transformation 
\begin{eqnarray}
\Phi  \to e^{-i{2\pi k/{3}}} \Phi, \quad
\Phi^{*} \to e^{i{2\pi k/{3}}}\Phi^{*}
\label{Z3}
\end{eqnarray}
with an arbitrary integer $k$, one can see that $\Omega$ has 
the RW periodicity~\cite{RW,Sakai,Kouno,Morita,Pagura_im}: 
\begin{eqnarray}
\Omega (\theta_u,\theta_d,\theta_s)=\Omega (\theta_u+2k\pi /3,\theta_d+2k\pi /3,\theta_s+2k\pi /3). 
\nonumber\\
\label{RW}
\end{eqnarray}
The RW periodicity does not mean that the system is 
$\mathbb{Z}_3$-symmetric, since 
the external parameters $\theta_f$ 
are shifted  by the $\mathbb{Z}_3$ transformation. 
An exception is the three degenerate flavor system with $\theta =2\pi/3$. 
In fact, the $\theta_f$ are shifted by the $\mathbb{Z}_3$ transformation, 
but the shifted $\theta_f$ are transformed back to the original by the 
relabeling of flavors, as mentioned in Sec. \ref{Introduction}. 
In the confinement phase appearing at low $T$ 
as a consequence of $\mathbb{Z}_3$ symmetry, 
the thermodynamic potential is obtained by setting $\Phi=0$ 
in \eqref{PNJL-Omega}. 
The resultant thermodynamic potential includes 
only three-quark configurations $e^{- 3 E_f/T}$. 
As a result of this property, the flavor symmetry broken 
by the flavor-dependent TBC is recovered 
in the confinement phase~\cite{Kouno_TBC,Sakai_TBC}. 

Finally we consider $\theta$-dependence of $\Phi$ at low $T$. 
In \eqref{PNJL-Omega}, $\Omega$ depends on $\Phi$ 
only through the Polyakov-loop potential ${\cal U}$ and 
the logarithmic term 
$F \equiv - T \sum_f (\ln{\cal F}_f+\ln{\cal F}_{\bar f})$. 
For small $T$, it is well satisfied that $T \ll M_f$ and $\Phi \ll 1$. 
Hence $F$ is approximated into 
\begin{eqnarray}
F \approx 
- 6 T \Phi N
\label{F-approx}
\end{eqnarray}
with 
\begin{eqnarray}
N=\sum_f e^{-\beta E^-_f},    
\label{F-approx-N}
\end{eqnarray}
where $\Phi$ and $N$ are real for any $\theta$, 
because ${\cal C}$ symmetry is preserved. 
For small $T$, the Polyakov-loop potential ${\cal U}$ has a global minimum 
at $\Phi=0$. The logarithmic term then moves the minimum point 
to positive (negative) $\Phi$, when $N$ is positive (negative). 
When $\theta$ varies from 0 to $2\pi/3$, $N$ changes the sign from plus 
to minus. Hence the minimum point moves from positive $\Phi$ 
to negative $\Phi$ as $\theta$ increases from 0 to $2\pi/3$. 
Eventually the confinement phase with $\Phi=0$ emerges at some value 
$\theta_{\rm conf}$ of $\theta$. 
It follows from $N=0$ that  
\begin{eqnarray}
\theta_{\rm conf} \approx \arccos{\left(-{1\over{2}}
e^{\beta (M_l-M_s)}\right)},   
\label{theta_solution}
\end{eqnarray}
and hence $\theta_{\rm conf}  =2\pi /3$ for three degenerate 
flavors ($M_l=M_s$) and $\pi /2$ for two flavors ($M_s=\infty$). 
This indicates that $\theta_{\rm conf}$ is between $2\pi /3$ and 
$\pi /2$ for 2+1 flavors, whereas 
$\theta_{\rm conf} = 2\pi /3$ for set S. 
Thus the confinement emerges, even if $\mathbb{Z}_3$ symmetry is not preserved.
The thermodynamic potential \eqref{PNJL-Omega} in the phase has 
only three-quark configurations $e^{- 3 E_f/T}$ for both 
of three degenerate flavors and 2+1 flavors. 
The only difference between the two cases is the value of $\theta_{\rm conf}$.

\section{Numerical results}
\label{results}

\subsection{The case of $\mu=0$}
\label{zero}

\begin{figure}[htbp]
\begin{center}
\includegraphics[width=0.3\textwidth]{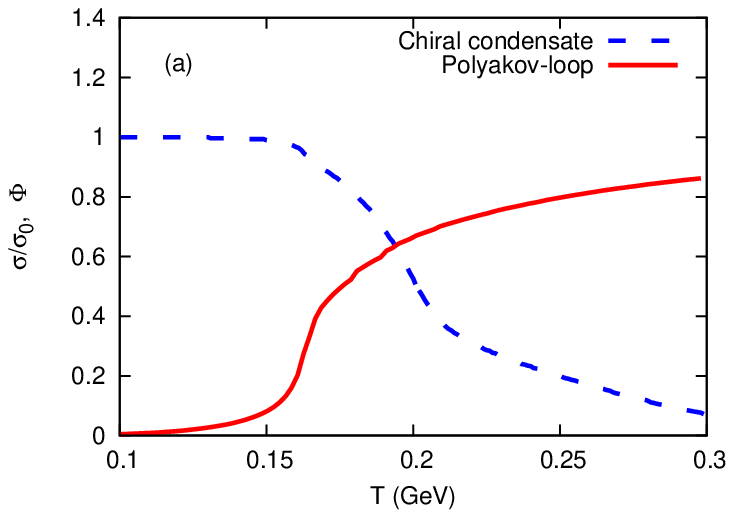}
\includegraphics[width=0.3\textwidth]{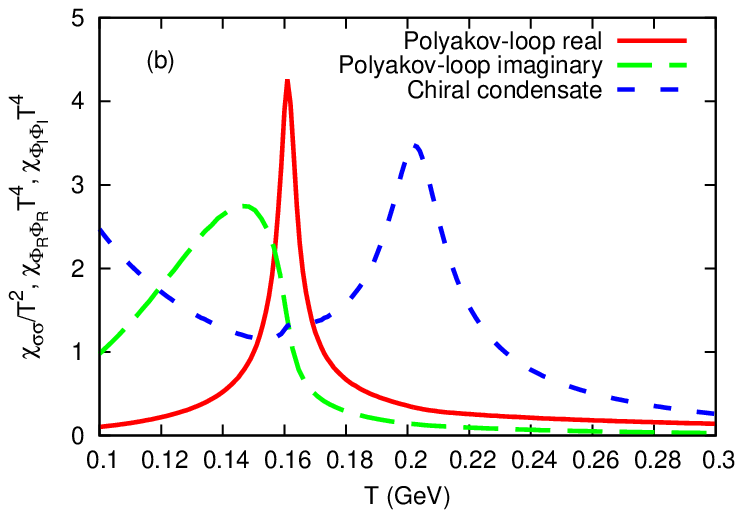}
\end{center}
\caption{$T$ dependence of (a) the chiral condensate $\sigma$ and 
the Polyakov loop $\Phi$ and (b) their susceptibilities $\chi_{\sigma\sigma}$ 
, $\chi_{\Phi_{\rm R}\Phi_{\rm R}}$ and $\chi_{\Phi_{\rm I}\Phi_{\rm I}}$ at $\mu_f =0$. 
Set R is taken in the PNJL calculation. 
The chiral condensate $\sigma$ is normalized by 
the value $\sigma_0$ at $T=0$. 
$\chi_{\Phi_{\rm R}\Phi_{\rm R}}$ and $\chi_{\Phi_{\rm I}\Phi_{\rm I}}$ are multiplied by 10 and 100, respectively.  
}
\label{mu0000_order}
\end{figure}

We first analyze the thermodynamics at $\mu=0$ through 
the PNJL calculation with set R. 
In panel (a) of Fig. \ref{mu0000_order}, $\sigma$ and $\Phi$ are 
plotted as a function of $T$.  
As $T$ increases, $\sigma$ decreases smoothly, 
while $\Phi$ increases continuously. 
Both the chiral and the deconfinement transition are thus crossover. 
Panel (b) corresponds to $T$ dependence of 
susceptibilities $\chi_{\sigma\sigma}$, $\chi_{\Phi_{\rm R}\Phi_{\rm R}}$ and $\chi_{\Phi_{\rm I}\Phi_{\rm I}}$.  
Here the pseudocritical temperature $T_{\rm C}$ ($T_{\rm D}$) 
of the chiral (deconfinement) transition is 
defined by the temperature at which 
$\chi_{\sigma\sigma}$ ($\chi_{\Phi_{\rm R}\Phi_{\rm R}}$) 
becomes maximum; in the present calculation, 
$T_{\rm C}=202$~MeV and $T_{\rm D}=161$~MeV. 
Since $\Phi=\Phi^*$, $\Phi_{\rm I}$ is zero. 
This means that $\Phi_{\rm I}$ itself is not a good order parameter of the deconfinement transition. 
The peak position of $\chi_{\Phi_{\rm I}\Phi_{\rm I}}$ 
does not coincide with that of $\chi_{\Phi_{\rm R}\Phi_{\rm R}}$, 
but it should be noted that 
$\chi_{\Phi_{\rm I}\Phi_{\rm I}}$ is rapidly changed at $T=T_{\rm D}$ where 
$\chi_{\Phi_{\rm R}\Phi_{\rm R}}$ has a peak. 


\subsection{The case of $\theta =2\pi /3$}

\begin{figure}[htbp]
\begin{center}
\includegraphics[width=0.3\textwidth]{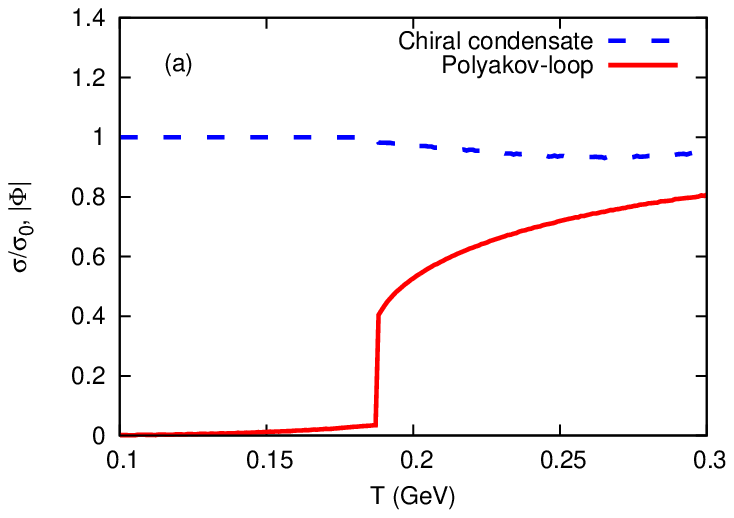}
\includegraphics[width=0.3\textwidth]{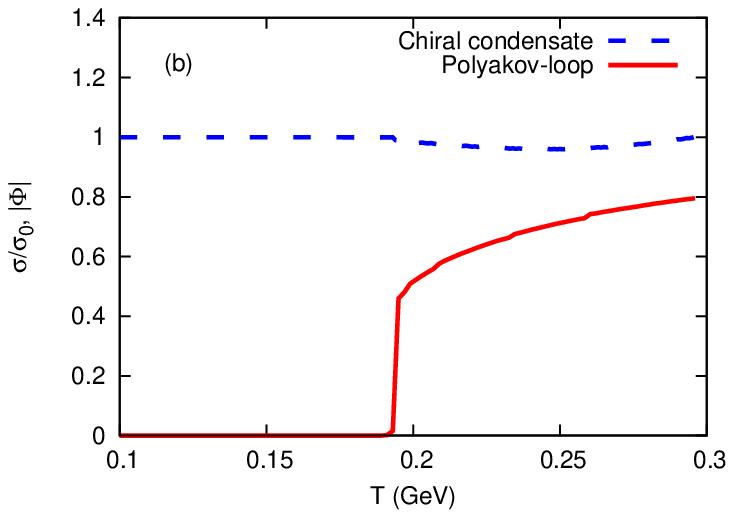}
\end{center}
\caption{$T$ dependence of the chiral condensate $\sigma$ and 
the absolute value $|\Phi|$ at $\theta =2\pi /3$. 
The PNJL calculation is done with set R in panel (a) and 
set S in panel (b). 
The chiral condensate $\sigma$ is normalized by 
the value $\sigma_0$ at $T=0$. 
}
\label{mu23pi_order}
\end{figure}

Next we analyze the thermodynamics at $\theta =2\pi /3$ through 
the PNJL model with set R and set S. 
First we consider set R. 
In Fig. \ref{mu23pi_order}(a), $\sigma$ and $|\Phi|$ are plotted as 
a function of $T$. 
The Polyakov loop $\Phi$ is real at small $T$ where ${\cal C}$ symmetry is preserved, 
but becomes complex at high $T$ where the ${\cal C}$ symmetry 
is spontaneously broken. The high $T$ region is called the RW phase; 
further discussion will be made in subsection \ref{Phase diagram} for the 
RW phase. 
As $T$ increases, $\Phi$ has a discontinuity at $T=T_{\rm D}=188$~MeV, whereas 
$\sigma$ decreases slowly. This indicates that 
the deconfinement transition is the first order. 
Thus the imaginary isospin chemical potential makes the deconfinement 
transition stronger. The same property is seen in the $N_f=2$ 
case~\cite{Sakai_imiso}.


\begin{figure}[htbp]
\begin{center}
\includegraphics[width=0.3\textwidth]{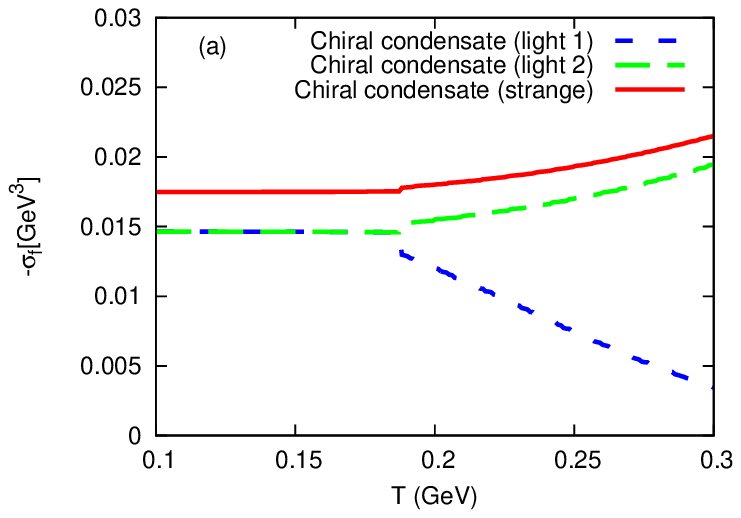}
\includegraphics[width=0.3\textwidth]{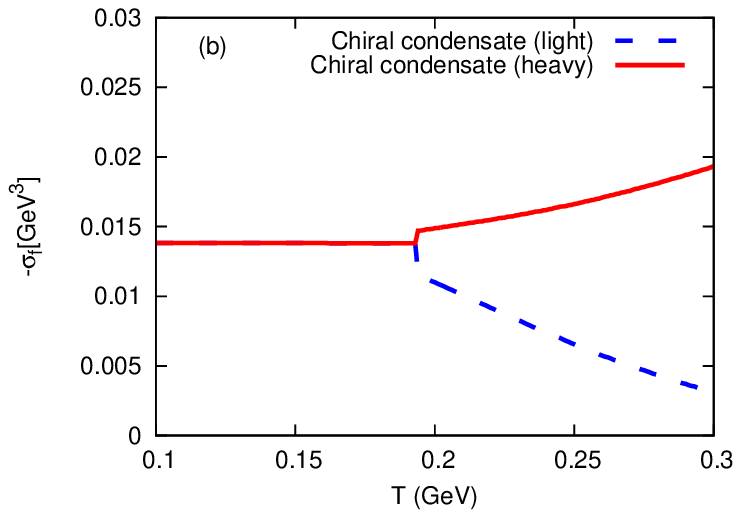}
\end{center}
\caption{$T$ dependence of the chiral condensates $\sigma_f$ 
at $\theta =2\pi /3$. 
The PNJL calculation is done with set R in panel (a) and 
set S in panel (b). 
In panel (a), the solid line represents s-quark, 
while the dashed and dotted lines correspond to light quarks.  
In panel (b), the solid and dashed lines agree with each other. 
}
\label{mu23pi_sigmaf}
\end{figure}

In the RW phase, the isospin symmetry between $u$ and $d$ is 
also broken due to the spontaneous breaking of ${\cal C}$ symmetry~\cite{Kouno}. 
Figure \ref{mu23pi_sigmaf}(a) shows $T$ dependence of $\sigma_f$ 
at $\theta =2\pi /3$ for the case of set R. 
As a consequence of the isospin symmetry breaking,
one of light quarks becomes heavier, while 
the other becomes lighter. 


Figure \ref{mu23pi_order}(b) is the same as Fig.~\ref{mu23pi_order}(a), 
but the PNJL calculation is done with set S. 
The order parameters $\sigma$ and $\Phi$ have 
similar $T$ dependence to those in Fig.~\ref{mu23pi_order}(a). 
However, $\Phi$ is zero 
below $T_{\rm D}=194$MeV, since ${\mathbb Z}_3$ symmetry is exactly preserved. 
${\mathbb Z}_3$ symmetry is spontaneously broken above $T_{\rm D}$. 
In this case, the confinement/decconfinement phase transition is governed 
by ${\mathbb Z}_3$ symmetry. 
Figure \ref{mu23pi_sigmaf}(b) is the same as Fig.~\ref{mu23pi_sigmaf}(a), 
but the PNJL calculation is done with set S. 
In panel (b), two of three quarks 
are always degenerate and become heavier at $T>T_{\rm D}$, 
while one of them becomes lighter there. 
The flavor symmetry broken by the TBC is found to be recovered 
in the confinement phase below $T_{\rm D}$.


\subsection{Phase diagram in the $\theta$-$T$ plane}
\label{Phase diagram}

\begin{figure}[htbp]
\begin{center}
\includegraphics[width=0.3\textwidth]{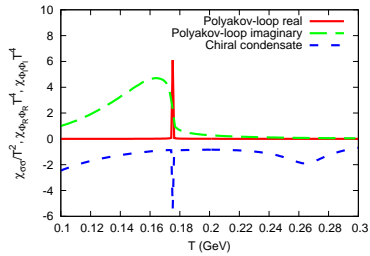}
\end{center}
\caption{$T$ dependence of $\chi_{\sigma\sigma}$, $\chi_{\Phi_{\rm R}\Phi_{\rm R}}$ and $\chi_{\Phi_{\rm I}\Phi_{\rm I}}$
at $\theta =\theta_{\rm CEP}$. 
The PNJL calculation is done with set R. 
$\chi_{\sigma\sigma}$ is multiplied by -1. 
$\chi_{\Phi_{\rm R}\Phi_{\rm R}}$ is divided by 10, while $\chi_{\Phi_{\rm I}\Phi_{\rm I}}$ is multiplied by 100. 
}
\label{mucrit_sus}
\end{figure}

Comparing Fig. \ref{mu0000_order} with Fig. \ref{mu23pi_order}, 
one can predict that there exists a critical endpoint (CEP) of 
deconfinement transition in the $\theta$-$T$ plane. 
In fact, the CEP appears 
at $\theta =\theta_{\rm CEP}=0.62\times {2\pi/3}=0.41\pi$, 
since the transition changes the order from crossover to first order there. 
Figure \ref{mucrit_sus} shows $T$ dependence of $\chi_{\Phi_{\rm R}\Phi_{\rm R}}$, $\chi_{\Phi_{\rm I}\Phi_{\rm I}}$ and $\chi_{\sigma\sigma}$ at $\theta =\theta_{\rm CEP}$.  
The susceptibility $\chi_{\Phi_{\rm R}\Phi_{\rm R}}$  diverges at 
$T=T_{\rm CEP}=175$~MeV.  
This indicates 
that the deconfinement phase transition is second order just on the CEP. 
Since $\Phi_{\rm R}$ and $\sigma$ are ${\cal C}$-even, they 
can be correlated with each other. 
In fact, $\chi_{\sigma\sigma}$ has a divergent peak on the CEP, 
although it has another peak at higher $T$.  
Meanwhile, $\Phi_{\rm I}$ is ${\cal C}$-odd, so that 
it is not divergent on the CEP, although it 
rapidly decreases there.


\begin{figure}[htbp]
\begin{center}
\includegraphics[width=0.3\textwidth]{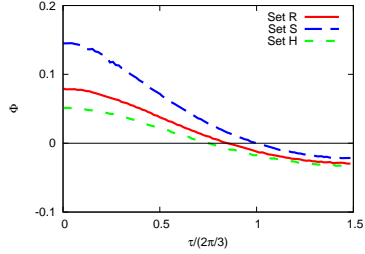}
\end{center}
\caption{$\theta$ dependence of $\Phi$ at $T=150$MeV. 
The PNJL calculations are done with three parameter sets of R, S and H. 
}
\label{Phi_theta-dep}
\end{figure}

As mentioned in Sec. \ref{PNJL model}, 
$\Phi$ becomes zero at some value $\theta_{\rm conf}$ of $\theta$ when 
$T$ is small, and $\theta_{\rm conf}$ is $2\pi /3$ 
for three degenerate flavors and less than $2\pi /3$ for 2+1 flavors. 
Figure \ref{Phi_theta-dep} shows $\theta$ dependence of $\Phi$ at low $T$. 
The PNJL calculation is done for three cases of set R, set S and set H. 
The value of $\theta_{\rm conf}$ is ${2\pi/3}$ for set S ($m_s=5.5$~MeV), 
$\sim 0.85\times {2\pi/3}=0.56\pi$ 
for set R ($m_s=140.7$~MeV) and $\sim 0.75\times {2\pi/3} ={\pi/2}$ for 
set H ($m_s=600$~MeV). The set-H case agrees with the two-flavor 
case~\cite{Sakai_imiso}. 
Strictly speaking, $\theta_{\rm conf}$ depends on $T$ for 2+1 flavors, 
but does not for three degenerate flavors, as shown in \eqref{F-approx-N}. 


\begin{figure}[htbp]
\begin{center}
\includegraphics[width=0.3\textwidth]{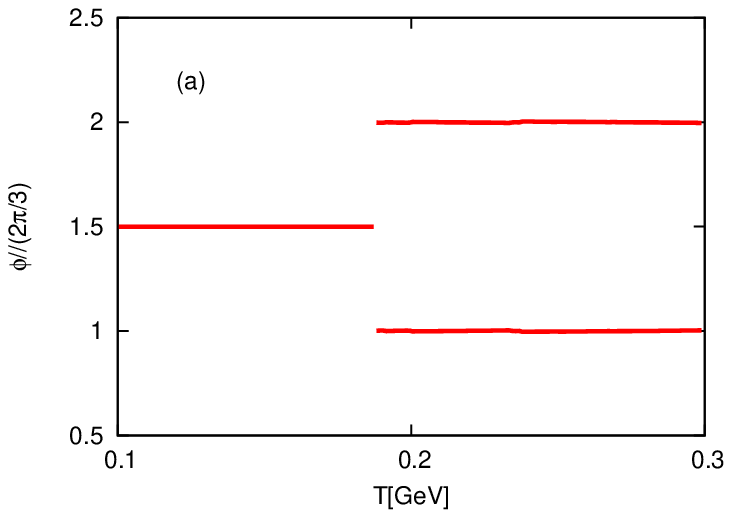}
\includegraphics[width=0.3\textwidth]{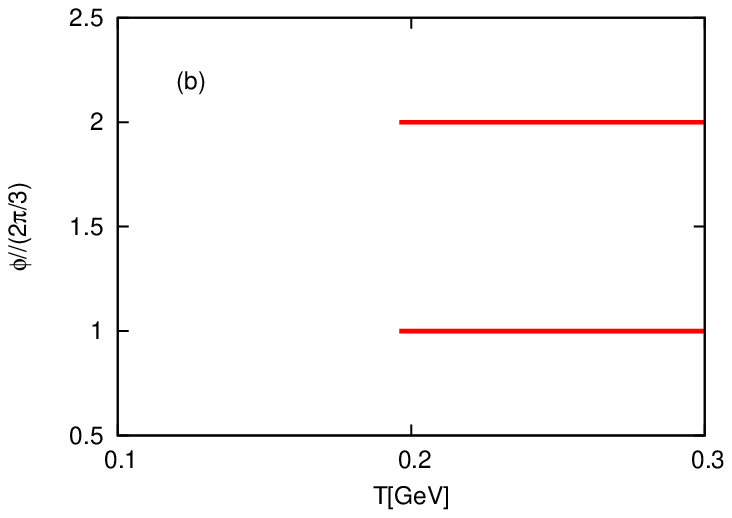}
\end{center}
\caption{$T$-dependence of the phase $\phi$ of $\Phi$ 
at $\theta =2\pi/3$.  
The PNJL calculation is done with (a) set R and (b) set S. 
Note that $\phi =4\pi /3=-2\pi/3$ mod $2\pi$. 
}
\label{sphase_T-dep}
\end{figure}

Figure \ref{sphase_T-dep} shows $T$dependence of 
the phase $\phi$ of $\Phi$ at $\theta =2\pi/3$.  
The phase $\phi$ is ${\cal C}$-odd and hence an 
order parameter of ${\cal C}$ symmetry. 
In panel (a) of set R, $\phi$ is $\pi$ at small $T$, and it 
jumps to about $\pm 2\pi/3$ at high $T$. This indicates 
that $\Phi$ is negative at small $T$ and ${\cal C}$ symmetry is broken 
at high $T$. 
Thus the system is in the RW phase at high $T$. 
Similar result is seen in panel (b) of set S. 
Here $\Phi$ is zero at small $T$ and hence $\phi$ is not defined there. 
At high $T$, $\phi$ is $\pm 2\pi/3$.  


\begin{figure}[htbp]
\begin{center}
\includegraphics[width=0.3\textwidth]{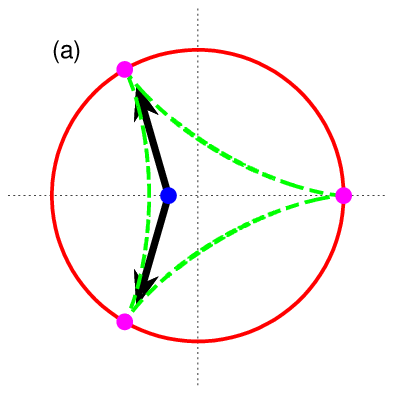}
\includegraphics[width=0.3\textwidth]{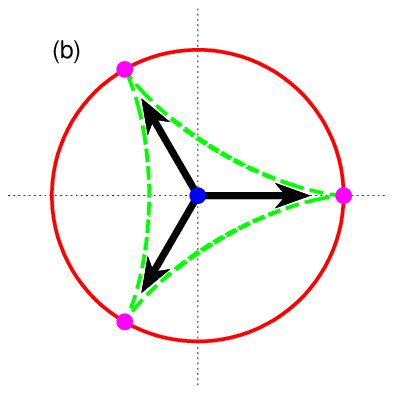}
\end{center}
\caption{Typical movement of $\Phi$ at $\theta =2\pi/3$ in (a) set R and (b) set S. 
The Polyakov loop $\Phi$ is not allowed to move outside the dashed lines, 
since the Polyakov-loop potential diverges on the dashed lines. 
}
\label{Phase-variation}
\end{figure}

$T$ dependence of $\phi$ at $\theta =2\pi/3$ shown above is illustrated in 
Fig. \ref{Phase-variation} as a movement of $\Phi$ 
with respect to increasing $T$ in the complex $\Phi$ plane. 
In panel (a) of set R, the RW phase transition as 
the spontaneous breaking of ${\cal C}$ symmetry 
is illustrated by two arrows from some negative value. 
The transition is mirror-symmetric about the 
real $\Phi$ axis in virtue of ${\cal C}$ symmetry. 
In panel (b) of set S, the deconfinement transition as 
the spontaneous breaking of $\mathbb{Z}_3$ symmetry is illustrated 
by three arrows. The transition is $\mathbb{Z}_3$-symmetric and a jump of 
$\Phi=0$ 
to $\Phi=\Phi', \Phi'e^{\pm 2\pi/3}$ for positive $\Phi'$ smaller than 1. 


\begin{figure}[htbp]
\begin{center}
\includegraphics[width=0.3\textwidth]{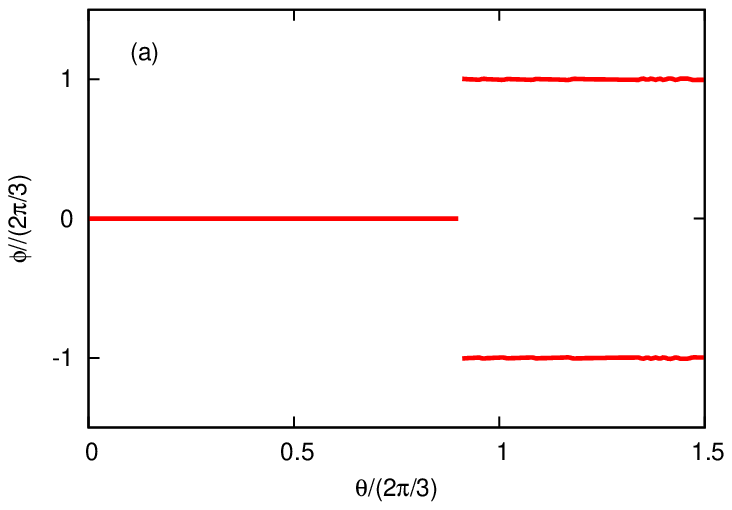}
\includegraphics[width=0.3\textwidth]{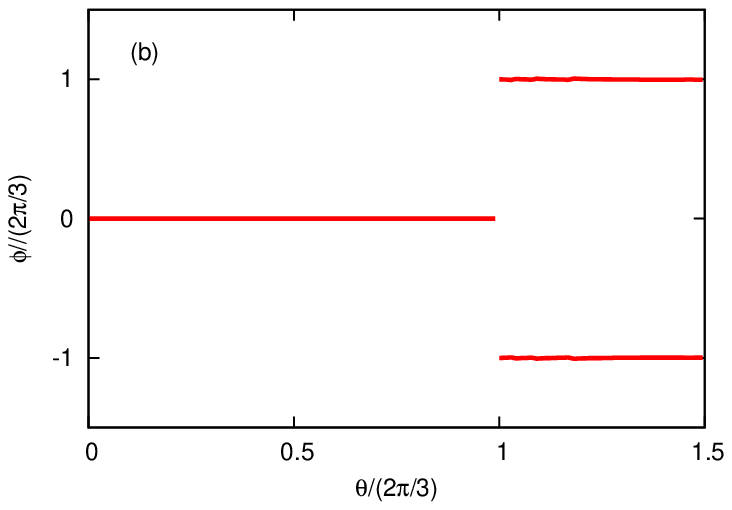}
\end{center}
\caption{$\theta$-dependence of the phase $\phi$ of $\Phi$ 
at $T=250$MeV.  
The PNJL calculation is done with (a) set R and (b) set S. 
}
\label{sphase_theta-dep}
\end{figure}

Figure \ref{sphase_theta-dep} shows $\theta$ dependence 
of $\phi$ at high $T$. 
The $\theta$ dependence is symmetric 
with respect to the line $\theta =\pi$, so we consider 
a range of $0 \le \theta \le \pi$ only. 
In panel (a) of set R, $\phi$ is zero at 
$0\le \theta < 0.91\times {2\pi/3}=0.61\pi$, 
but diverges into two solutions $\phi\approx \pm 2\pi /3$ 
at $\theta > 0.61\pi$.  The system is in the RW phase 
when $\theta > 0.61\pi$; here one solution is a ${\cal C}$ image of 
the other solution. 
Similar result is seen in panel (b) of set S; 
$\phi$ is zero at $0\le \theta< 2\pi /3$, while 
$\phi \approx \pm 2\pi /3 $ at $\theta >2\pi /3$.

The RW phase transition was discovered first 
in the $\theta_q$-$T$ plane \cite{RW}, 
where $\theta_q$ is the dimensionless imaginary quark-number 
chemical potential defined by the quark-number chemical potential $\mu_q$ 
as $\mu_q=iT \theta_q$. 
The quark-number density is discontinuous at $\theta_q=\pi/3$ mod $2\pi/3$, 
when $T$ is higher than some temperature $T_{\rm RW}$. 
This first-order phase transition is now called 
the RW phase transition. 
It was found in Ref. \cite{Kouno} that 
${\cal C}$ symmetry is also spontaneously broken 
in the RW phase transition. 
One can then define the RW phase transition 
by the spontaneous breaking of ${\cal C}$ symmetry. 
In the $\theta_q$-$\theta$-$T$ space, the RW phase appears as 
a plane of $\theta_q=0$, $\theta_0 \le  \theta \le \pi$ and 
$T > T_{\rm RW}$. Here $\theta_0$ is a critical value of $\theta$ and 
depends on $T$ for set R but not for set S; 
for example, $\theta_0=0.93 \times 2\pi/3$ at $T=250$ MeV for set R and 
$2\pi /3$ at $T>T_{\rm RW}$ for set S. 


\begin{figure}[htbp]
\begin{center}
\includegraphics[width=0.3\textwidth,angle=0]{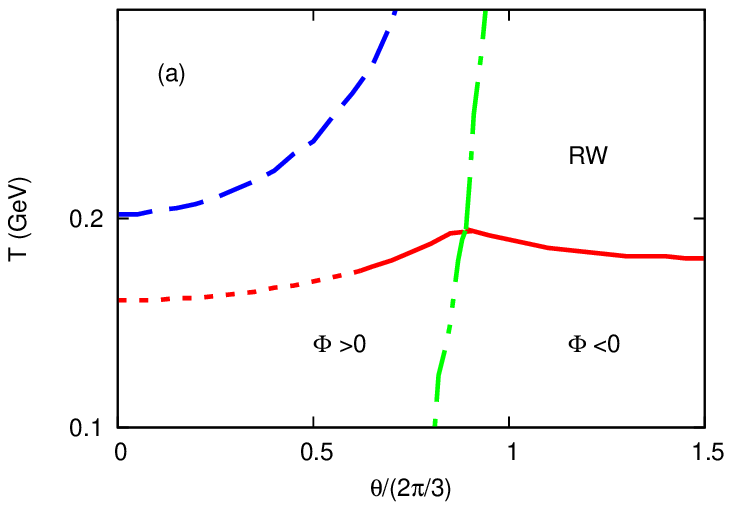}
\includegraphics[width=0.3\textwidth,angle=0]{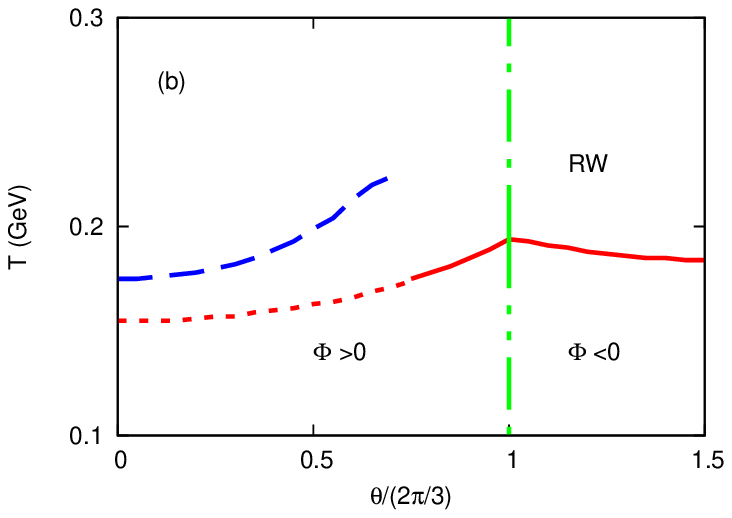}
\includegraphics[width=0.3\textwidth,angle=0]{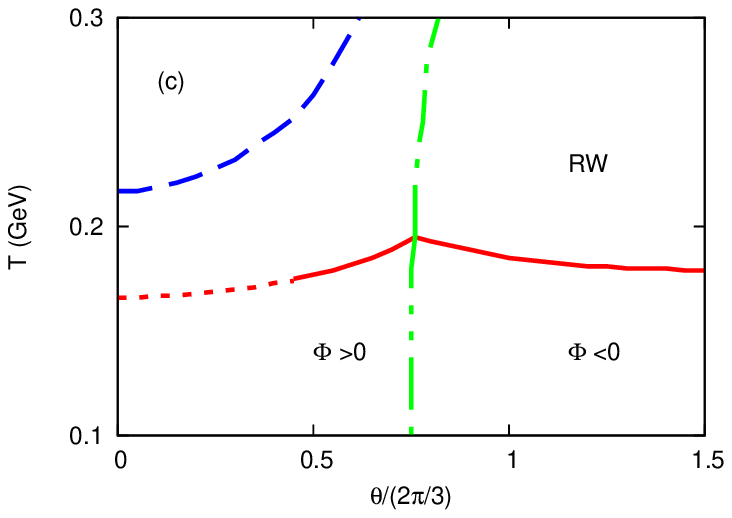}
\end{center}
\caption{Phase diagram in the $\theta$-$T$ plane. 
The PNJL calculation is done with (a) set R, 
(b) set S and (c) set H. 
The solid, dotted, dashed lines represent first-order deconfinement, 
crossover deconfinement and crossover chiral transitions, respectively. 
The dot-dashed line is the left boundary of the RW phase, 
while the lower boundary of the RW phase is 
the solid line. 
The dot-dot-dashed line is the line on which $\Phi =0$. 
Below the RW region, $\Phi$ becomes negative.  
}
\label{phase}
\end{figure}

Figure \ref{phase} shows the phase diagram in the $\theta$-$T$ plane. 
Three panels correspond to parameter sets of R, S and H, respectively. 
In panel (a) of set R, as mentioned above, 
there is a CEP of deconfinement transition 
at $\theta_{\rm CEP}=0.41\pi $ and $T_{\rm CEP}=175$~MeV. 
Meanwhile, the chiral transition is crossover. 
Phase diagrams for set S and set H have almost the same structure as that for 
set R, as shown in panels (b) and (c). 
The dot-dashed line (the left boundary  of RW phase) and 
the dot-dot-dashed line (the line of $\Phi =0$) are shifted to the left, 
when $m_s$ becomes heavy.  
The shift of the dot-dot-dashed line indicates that 
the confinement at $\theta =0$ becomes stronger as $m_s$ increases. 
In all the cases, the chiral restoration is weakened at 
large $\theta$, so the chiral-transition lines are not shown there.  


\begin{figure}[htbp]
\begin{center}
\includegraphics[width=0.3\textwidth,bb=50 60 210 160,clip]{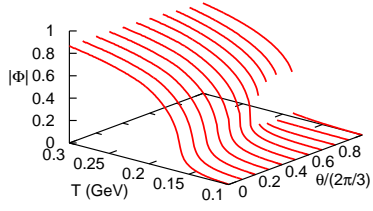}
\end{center}
\caption{The absolute value of $\Phi$ as a function of $\theta$ and $T$. 
The PNJL calculation is done with set R. 
}
\label{phase_PNJL}
\end{figure}

In Fig.~\ref{phase_PNJL}, $|\Phi |$ is plotted as a function of $\theta$ and 
$T$. For large $\theta$, $|\Phi|$ has an abrupt jump as $T$ increases. 
This jump  indicates that the deconfinement transition 
is first order. As $\theta$ decreases, the order of the deconfinement 
transition is changed into crossover. 
The crossover deconfinement transition at $\theta=0$ is thus 
a remnant of the first-order deconfinement transition at large $\theta$. 


\subsection{Susceptibilities at zero and finite $\theta$}

\begin{figure}[htbp]
\begin{center}
\includegraphics[width=0.3\textwidth]{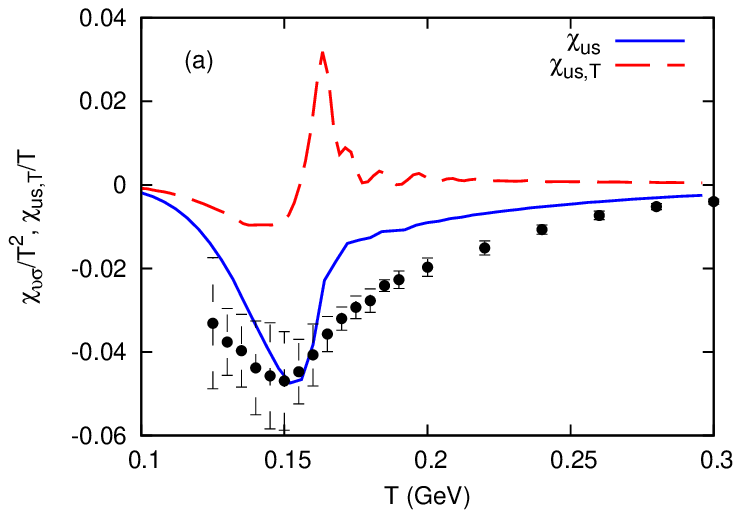}
\includegraphics[width=0.3\textwidth]{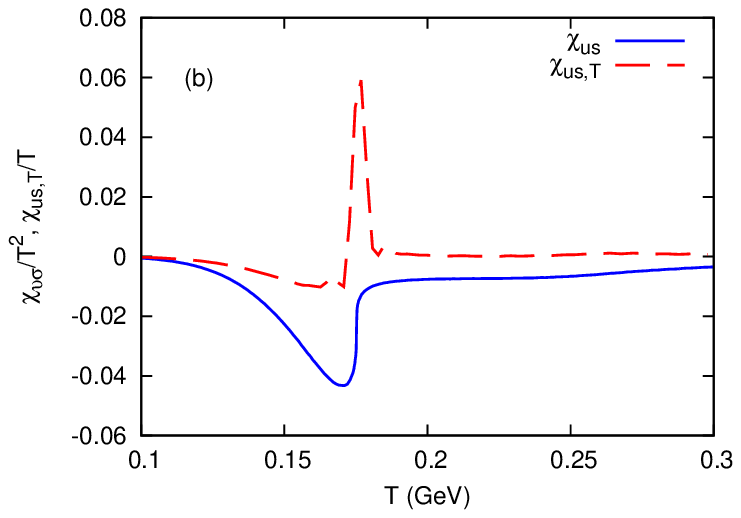}
\end{center}
\caption{
$T$ dependence of quark number susceptibility $\chi_{us}$ (solid line) 
and its derivative (dashed line) with respect to $T$ 
for (a) $\theta =0$ and (b) $\theta=\theta_{\rm CEP}$. 
The PNJL calculation is done with set R, and the resultant 
$\chi_{us}$ is multiplied by 20. 
The corresponding LQCD result~\cite{BFKKRS}, 
shown by dots with errorbars in panel (a), are not normalized. 
}
\label{number_sus}
\end{figure}

Figure \ref{number_sus}(a) shows $T$ dependence of 
the non-diagonal element $\chi_{us}$ 
of quark number density susceptibilities and its derivative $\chi_{us,T}$ 
with respect to $T$ for the case of $\theta =0$. 
As an interesting property, $\chi_{us}$ suddenly changes 
near $T_{\rm D}=161$~MeV, and consequently $|\chi_{us,T}|$ has a peak there. 
Comparing $\chi_{us}$ 
in Fig. \ref{number_sus}(a)
with $\chi_{\Phi_{\rm I}\Phi_{\rm I}}$ in Fig. \ref{mu0000_order}(b) shows that that the two quantities are strongly correlated with each other. 

The interplay between $\chi_{us}$ and $\chi_{\Phi_{\rm I}\Phi_{\rm I}}$ 
can be understood as follows. 
The stationary condition 
\begin{eqnarray}
{\partial \over{\partial \varphi_i}}\Omega (T,\mu,\varphi )=0
\label{statinary}
\end{eqnarray}
and its derivatives
\begin{eqnarray}
{D \over{D \mu_f}}\left[{\partial \over{\partial \varphi_i}}\Omega (T,\mu,\varphi (T,\mu ))\right]=0, 
\label{statinary_derivative}
\end{eqnarray}
lead to a relation between $\chi_{ff^\prime}$ 
and $\chi_{\varphi_i\varphi_j}$ as  
\begin{eqnarray}
\chi_{ff^\prime}&=&-{D^2\Omega \over{D\mu_fD\mu_{f^\prime}}}
\nonumber\\
&=&-{\partial^2 \Omega (T,\mu, \varphi )\over{\partial\mu_f\partial \mu_{f^\prime}}}
\nonumber\\
&&+{\partial^2 \Omega (T,\mu,\varphi )\over{\partial\mu_f\partial\varphi_i}}\chi_{\varphi_i\varphi_j}
{\partial^2 \Omega (T,\mu,\varphi )\over{\partial\mu_{f^\prime}\partial\varphi_j}}. 
\label{sus_relation}
\end{eqnarray}
For later convenience, we define $\partial_X=\partial/\partial X$. 
As for $\chi_{us}$,  the second derivative 
$\partial_{\mu_u}\partial_{\mu_s} \Omega$ vanishes at $\mu_f=0$, 
since $\Omega$ is $\mu_f$-even for each $f$ and hence 
$\partial_{\mu_u}\partial_{\mu_s} \Omega$ is $\mu_f$-odd. 
Similarly, $\mu_f$-odd quantities $\partial_{\mu_f} \partial_{\sigma} \Omega$ 
, $\partial_{\mu_f} \partial_{\sigma'} \Omega$ and 
$\partial_{\mu_f} \partial_{\Phi_{\rm R}}\Omega$ vanish at $\mu_f=0$, whereas 
a $\mu_f$-even quantitiy $\partial_{\mu_f} \partial_{\Phi_{\rm I}} \Omega$ 
does not. These properties lead to 
\begin{eqnarray}
\chi_{us}&=&
{\partial^2 \Omega (T,\mu_u,\varphi )\over{\partial\mu_u\partial\Phi_{\rm I}}}\chi_{\Phi_{\rm I}\Phi_{\rm I}}
{\partial^2 \Omega (T,\mu_u,\varphi )\over{\partial\mu_s\partial\Phi_{\rm I}}}
\label{sus_relation_us}
\end{eqnarray}
Thus $\chi_{us}$ is correlated with not 
$\chi_{\Phi_{\rm R}\Phi_{\rm R}}$ but $\chi_{\Phi_{\rm I}\Phi_{\rm I}}$. 
The PNJL result shown in Fig. \ref{number_sus}(a) well simulates $T$ dependence of $\chi_{us}$ calculated with 
LQCD~\cite{BFKKRS,HotQCD}, 
although the former underestimates the latter for the magnitude 
of $\chi_{us}$. 
This underestimation may stem from the fact 
that the present model does not treat baryon degrees of freedom explicitly. 


Similar behavior is seen on the CEP, as shown in Fig. \ref{number_sus}(b).  
The relation (\ref{sus_relation_us}) persists for finite $\theta$, 
since ${\cal C}$ symmetry is preserved there. 
Therefore $|\chi_{us}|$ rapidly decreases at $T=T_{\rm CEP}$, and consequently $\chi_{us,T}$ has a peak there. 
The peak is even more conspicuous for $\theta =\theta_{\rm CEP}$ 
than for $\theta=0$. 


\section{Summary}
\label{Summary}

We have investigated the confinement mechanism in three-flavor QCD 
with finite imaginary isospin chemical potentials 
$(\mu_u,\mu_d,\mu_s)=(i\theta T,-i\theta T,0)$, using the PNJL model. 
Two cases of three degenerate flavors and 2+1 flavors were taken.   
In the former case, the system has $\mathbb{Z}_{3}$ symmetry at 
$\theta=2\pi/3$. The symmetry is not spontaneously broken 
for low $T$ and hence 
the Polyakov loop $\Phi$ is zero there. 
In the latter case, $\mathbb{Z}_{3}$ symmetry is 
explicitly broken for any $\theta$, but $\Phi$ becomes zero at 
$\theta=\theta_{\rm conf}$ smaller than $2\pi/3$. 
Thus the confinement phase defined by 
$\Phi=0$ is realized, even if the system does not have $\mathbb{Z}_{3}$ 
symmetry exactly.  
In the confinement phase, the static-quark free energy, $-T \ln[\Phi]$, 
diverges. 
This means that one can consider the confinement by using $\Phi$ 
and regard it as an order parameter of the confinement/deconfinement 
transition, even if $\mathbb{Z}_{3}$ symmetry is not preserved exactly.

The phase diagram is determined in the $\theta$-$T$ plane. 
There is a CEP of deconfinement transition at a finite value 
$\theta_{\rm CEP}$ of $\theta$. 
This is a good contrast with the CEP of chiral transition at 
real quark-number chemical potential. 
As another interesting point, there is a line of $\Phi =0$ for both cases of 
three degenerate flavors and 2+1 flavors. On the line, the confinement phase 
is realized and the thermodynamic potential has only three-quark 
configurations in which red, green and blue quarks are 
statistically in the same state. This property is independent of 
whether the system has $\mathbb{Z}_{3}$ symmetry or not.

In the $\theta$-$T$ plane, the confinement transition is crossover on 
the axis of $\theta=0$, while it is first order on the line of $\Phi=0$. 
This means that the deconfinement crossover at $\theta=0$ is a remnant 
of the first-order deconfinement transition on the line of $\Phi=0$. 
Hence the distance of the line of $\Phi=0$ 
from the axis of $\theta=0$ shows how strong the confinement property is 
at $\theta=0$. The line of $\Phi=0$ is moved to smaller $\theta$, 
as $m_s$ increases. This means that the confinement property is stronger 
in the two-flavor case than in the 2+1 flavor case. 
This statement 
may be consistent with LQCD results~\cite{Sasaki-T_Nf3,Borsanyi,Bazavov}. 
The non-diagonal element $\chi_{us}$ of quark number susceptibilities 
is correlated with not $\chi_{\Phi_{\rm R}\Phi_{\rm R}}$ 
but $\chi_{\Phi_{\rm I}\Phi_{\rm I}}$ 
for zero and finite $\theta$. Hence 
$|\chi_{us}|$ does not have a peak 
near the pseudocritical temperature defined by $\chi_{\Phi_{\rm R}\Phi_{\rm R}}$, 
but rapidly changes there. 
As a consequence of this property, its derivative $\chi_{us,T}$ 
has a peak there. 
The peak is more conspicuous for finite $\theta$ than for zero $\mu$.

The present results are derived with the PNJL model, but 
these can be checked by LQCD simulations directly, 
since the simulations are free from the sign problem for 
imaginary isospin chemical potential.

\noindent
\begin{acknowledgments}
The authors thank A. Nakamura, T. Saito, K. Nagata and K. Kashiwa for useful discussions. 
H.K. also thanks M. Imachi, H. Yoneyama, H. Aoki and M. Tachibana for useful discussions. 
T.S. is supported by JSPS KAKENHI Grant Number 23-2790. 
Y.S. is supported by RIKEN Special Postdoctoral Researchers Program.
\end{acknowledgments}


\end{document}